\documentstyle[11pt,fleqn]{article}
\newcommand{\preprint}[1]{\hfill{\sl preprint - #1}\par\bigskip\par}
\def\title{\par\bigskip\begin{center}\bf\LARGE}
\def\endtitle{\end{center}\par\bigskip\par\rm\normalsize}
\def\instit{\begin{center}\large}
\def\endinstit{\end{center}\rm\normalsize}
\def\references{\begin{thebibliography}{10}}
\def\endreferences{\end{thebibliography}\end{document}}

\newcommand{\s}[1]{\section{#1}\renewcommand{\theequation}
        {\mbox{\arabic{section}.\arabic{equation}}}\setcounter{equation}{0}}

\renewcommand{\author}[1]{\begin{center}\Large #1\end{center}}
\renewcommand{\date}[1]{\par\bigskip\par\sl\hfill #1\par\medskip\par}
\newcommand{\email}[1]{e-mail: \sl #1@itnvax.science.unitn.it,
                               #1@itncisca.bitnet, 37953::#1}
\newcommand{\femail}[1]{\footnote{\email{#1}}}
\newcommand{\pacs}[1]{\smallskip\noindent{\sl PACS number(s):
                       \hspace{0.3cm}#1}\par\bigskip}
\newcommand{\babs}{\hrule\par\begin{description}\item{Abstract: }\it}
\newcommand{\eabs}{\par\end{description}\hrule\par\medskip\rm}
\newcommand{\ack}[1]{\par\section*{Acknowledgments} #1}
\newcommand{\M}{{\cal M}}              
\newcommand{\ca}[1]{{\cal #1}}         
\newcommand{\hs}{\qquad\qquad}         
\newcommand{\nn}{\nonumber}            
\newcommand{\beq}{\begin{eqnarray}}    
\newcommand{\eeq}{\end{eqnarray}}      
\newcommand{\beqn}{\begin{eqnarray}}   
\newcommand{\eeqn}{\end{eqnarray}}     
\newcommand{\at}{\left(}               
\newcommand{\aq}{\left[}               
\newcommand{\ct}{\right)}              
\newcommand{\cq}{\right]}              
\newcommand{\C}{\mbox{I$\!\!\!$C}}               
\newcommand{\ii}{\infty}                         
\newcommand{\X}{\times}                          
\newcommand{\fr}[2]{\mbox{$\frac{#1}{#2}$}}      
\newcommand{\Tr}{\,\mbox{Tr}\,}                  
\newcommand{\Res}{\,\mbox{Res}}                
\renewcommand{\Re}{\,\mbox{Re}\,}                
\newcommand{\lap}{\triangle}                     

\newcommand{\be}{\beta}
\newcommand{\ga}{\gamma}

\newcommand{\ep}{\varepsilon}
\newcommand{\ze}{\zeta}

\newcommand{\ka}{\kappa}
\newcommand{\la}{\lambda}

\newcommand{\om}{\omega}

\newcommand{\Ga}{\Gamma}

\newcommand{\hphi}{\hat{\phi}}
\newcommand{\tem}{\sum_{n=-\infty}^{\infty}}
\newcommand{\vol}{V_S}
\newcommand{\veff}{V_{eff}(\hphi)}
\newcommand{\vone}{V_T^{(1)}(\hphi)}
\newcommand{\ablt}{\frac{\partial^2}{\partial\tau^2}}
\begin{document}

\preprint{UTF 301}

\begin{title}
Finite temperature effective potential\\
on hyperbolic spacetimes\end{title}
\author{Guido Cognola\femail{cognola}}
\begin{instit}
Dipartimento di Fisica, Universit\`a di Trento \\
and
Istituto Nazionale di Fisica Nucleare \\
Gruppo Collegato di Trento, Italia
\end{instit}
\author{Klaus Kirsten\footnote{
Address after October 1, 1993: Department E.C.M., Faculty of
Physics, Barcelona University, Diagonal 647, E-08028 Barcelona, Spain}
}
\begin{instit}
Dipartimento di Fisica, Universit\`a di Trento, Italia
\end{instit}
\author{Luciano Vanzo\femail{vanzo}, Sergio Zerbini\femail{zerbini}}
\begin{instit}
Dipartimento di Fisica, Universit\`a di Trento \\
and Istituto Nazionale di Fisica Nucleare \\
Gruppo Collegato di Trento, Italia
\end{instit}

\babs
The finite temperature one-loop effective potential for a scalar field
defined on an ultrastatic spacetime, whose spatial part is a compact
hyperbolic manifold, is studied. Different analytic expressions,
especially valuable at low and high temperature are derived. Based on
these results, the symmetry breaking and the topological mass
generation are discussed.
\eabs

\s{Introduction}
One major area of research in quantum field theory in curved spacetime
in the 1980s has been on interacting quantum fields and their
implications (\cite{BD82,brandenberger85}, for a review see
\cite{BOS92}). The central object of physical
interest is the effective potential which should define the dynamics
of the early universe at scales bigger than the Planck scale. In
particular in new inflationary models
\cite{AS82,L82} the effective cosmological
constant is obtained from an effective potential which includes
quantum corrections to the classical potential of a scalar field
\cite{CW73}.

Unfortunately there are a lot of difficulties in calculating the
effective potential or action in a quantum field theory in a general
curved spacetime \cite{BOS92}. Only
approximations like the derivative expansion in the background field
at zero \cite{oconnorhu84,guven87,CKV93unv} and at
finite temperature
\cite{critchleyhustylianopoulos87,kirstenself} are
available.

Therefore it is very natural to deal with some specific spaces which
are interesting from the cosmological viewpoint. For example the role
of constant curvature \cite{OHS83} and of anisotropy in
different Bianchi type universes
\cite{OHS85,oconnorhu86,F84,huang90,R87,R87a,hartlehu79,berkin92}
has been considered in detail.

Furthermore the importance of nontrivial topology (imposed, for
example, by finite temperature or compactified spatial sections) has
been emphasized at several places. Apart from its well know influence
on the effective mass of the field
\cite{DJ74,A90a,FY79,ford80,T80,T80c,KN81,W74,L79,G82,DS81,DS80,ER90a,EKK93unv}
and consequently also on particle creation
\cite{L83,G84}, let us mention as a motivation for such
considerations that possibly the universe as a whole exhibits
nontrivial topology
(\cite{ellis71,ellisschreiber86,fagundes93} and references
therein).

Most of the mentioned literature has been concerned with only one
compact dimension (representing imaginary time or a compact spatial
dimension) and with a spacetime of vanishing curvature. However, in a
cosmological setting the inclusion of curvature is of course
necessary. For that reason, in a previous paper
\cite{CKZ93} we started recently
investigations on the influence of nontrivial topology together with
nonvanishing curvature on the effective potential in a
self-interacting scalar field theory. In that article we restricted
ourselves to zero temperature. However, in the early universe the
inclusion of nonvanishing temperature is necessary and this is the aim
of the following article, which is organized as follows.

In Sec.~2 we remind the reader very briefly of the definition of
the effective potential in a self-interacting scalar field theory.
Afterwards different techniques are applied to find analytic results
for the potential. First, in Sec.~3, the Selberg trace formula is
used to derive a very compact form of the effective potential of the
theory. Then, in Sec.~4, other representations of such a result, but
especially useful at low and high temperature, are given.
For both results the physical implications in terms of the
mass of the field are discussed in Sec.~5. The main
results are summarized in the conclusions.

\s{One-loop effective potential of the self-interacting theory}
The concept of the effective action and the effective potential is
well discussed in the literature
\cite{brandenberger85,BOS89,CW73,jackiw74}
and the introduction of these quantities will be very brief.

We consider a self-interacting scalar field coupled to gravity on a
four-dimensional spacetime with a compact hyperbolic spatial section
$H^3/\Ga$, $H^3$ being the Lobachevsky space and $\Gamma$
a discrete group of isometries of $H^3$,
which is assumed to have only hyperbolic elements, so the Riemannian
structure is smooth.
The theory at finite temperature $T=1/\be$ may be obtained by
compactifying the imaginary time $\tau$ and assuming
the field $\phi(\tau,\vec x)$ to fulfill periodic boundary
conditions with respect to $\tau$, this means
$\phi(\tau,\vec x)=\phi(\tau+\be,\vec x)$.
In this way, the manifold becomes an Euclidean space
of the form ${\cal M}=S^1\times H^3/\Gamma$, $S^1$ being a
circumference.

The theory is described by the action
\beq
S(\phi)=\int_0^{\beta}\,d\tau\,\int_{H^3/\Ga}
\left[-\frac12\phi\left(\frac{\partial^2}{\partial\tau^2}
+\Delta_S\right)\phi+V(\phi)\right]\,|g|^{\frac12}\,d^3x\:,
\label{21}
\eeq
with the Laplace-Beltrami operator $\Delta_S$ acting on the spatial
section $H^3/\Gamma$ of ${\cal M}$ and the
classical potential
\beq
V(\phi)=\frac{m^2\phi^2}2+\frac{\xi R\phi^2}2
+\frac{\lambda\phi^4}{24}\:.
\eeq
As usual $R$ is the scalar curvature of $\M$, while $\xi$ and
$\la$ are arbitrary dimensionless parameters (coupling constants).

The action (\ref{21}) has a minimum at $\phi =\hphi$ which satisfies
the classical equation of motion
\beq
-\left(\ablt+\Delta_S\right)\hphi+V'(\hphi)=0\:.
\label{22}
\eeq
Quantum fluctuations $\phi'=\phi-\hphi$ around the classical
background $\hphi$ satisfy to lowest order in $\phi'$
an equation of the form
\beq
A\phi'(x)=\left(-\ablt-\Delta_S+V''(\hphi )\right)\phi'(x)
=\left(-\ablt+L\right)\phi'(x)=0\:,
\label{23}
\eeq
where we have introduced the operator $L=-\lap_S+V''(\hphi)$.

Assuming a constant background field $\hphi$ the concept of the
effective potential is well defined \cite{berkin92}.
Formally it is given by
\beq
\veff=V(\hphi)+\hbar\vone+{\cal O}(\hbar^2)\:,
\label{24}
\eeq
with the purely classical part $V(\hphi)$ and with the one-loop
quantum corrections
\beq
V^{(1)}(\hphi)=\frac1{2\beta\vol}\ln\det(A\ell^2)
=V_0^{(1)}(\hphi)+\vone\:,
\label{25}
\eeq
where $\vol$ is the volume of the spatial hyperbolic part and $\ell$
a length renormalization scaling parameter.
Moreover, we distinguished between zero ( $V_0^{(1)}(\hphi)$)
and finite ($\vone$) temperature contributions.
Using $\ze$-function prescription for the regularization of
functional determinants \cite{hawking77,critchleydowker76},
Eq.~(\ref{25}) assumes the form
\beq
V^{(1)}(\hphi)=-\frac1{2\beta\vol}\left[\zeta'(0|A)
+\zeta(0|A)\ln\ell^2\right]\:,
\label{26}
\eeq
where $\zeta(s|A)$ is the $\ze$-function associated with the operator
$A$ and $\zeta'(s|A)$ its derivative with respect to $s$.
More explicitly, using the Poisson summation formula, one has
\beq
\zeta(s|A)&=&\tem\sum_j\left[\om_j^2+(2\pi n/\beta)^2\right]^{-s}
=\tem\ze(s|L+[2\pi n/\beta]^2)\label{27a}\\
&=&\frac{\be}{\pi}\tem \int_0^{\ii}\ze(s|L+t^2)\,\cos n\be t \,dt
\:,\label{27}
\eeq
$\om^2_j$ being the eigenvalues of the operator $L$.
The term $n=0$ in Eq.~(\ref{27a}) leads to the function
$\zeta(s|L)=\sum_j\left(\om_j^2\right)^{-s}$ on the
spatial section.
In the explicit computation, it is convenient to distinguish between
$n=0$ and $n\neq0$. Then,
using Eq.~(\ref{27}) and the Mellin
representation of $\ze$-function, we obtain
\beq
\zeta(s|A)
-\frac{\be\Ga(s-\fr{1}{2})\ze(s-\fr{1}{2}|L)}
{\sqrt\pi\Ga(s)}&=&\frac{2\be}{\pi}
\sum_{n=1}^\ii\int_0^{\ii}\ze(s|L+t^2)\,\cos n\be t\,dt\:,
\label{z2}\\
&=&\frac{\be}{\sqrt{\pi}\Ga(s)}
\sum_{n=1}^\ii\int_0^{\ii}t^{s-3/2}
e^{-n^2\be^2/4t}\Tr e^{-tL}\,dt
\:.\label{z3}
\eeq
We note that the $n=0$ term in Eqs.~(\ref{z3}) and (\ref{z2})
gives the zero temperature contribution $V_0^{(1)}(\hphi)$
to the effective potential.
Such a contribution, which requires renormalization, has been
extensively considered in Ref.~\cite{CKZ93}
to which we refer the reader for a detailed discussion.
Here we shall concentrate our attention only on the effects
due to non zero temperature.
Using Eqs.~(\ref{26}), (\ref{z2}) and (\ref{z3}) we get
\beq
\vone&=&-\frac1{\pi\vol}\,\lim_{s\to0}\,
\sum_{n=1}^{\ii}\int_{0}^{\ii}\ze'(s|L+t^2)\,\cos n\be t\,dt\:,
\label{v2}\\
&=&-\frac{1}{\sqrt{4\pi}\vol}
\sum_{n=1}^\ii\int_0^{\ii}t^{-3/2}
e^{-n^2\be^2/4t}\Tr e^{-tL}\,dt\:,
\label{v3}\\
&=&\frac{1}{\be\vol}\Tr\ln\at
1-e^{-\be Q}\ct\:,
\label{v4}
\eeq
where the operator $Q=|L|^{1/2}$ has eigenvalues $\om_j$.
As it is well known, low temperature contributions to the
effective potential are exponentially vanishing.

In view of a high temperature expansion,
it is convenient to write Eq.~(\ref{v3}) in a more useful form.
To this aim we recall the Mellin transforms
\beq
\hat f(z)=\int_0^\ii x^{z-1} f(x) \,dx\:,
\eeq
\beq
f(x)=\frac1{2\pi i}
\int_{\Re z=c} x^{-z}\hat f(z)\,dz\:,
\eeq
$c$ being a real number belonging to the strip in which
$\hat f(z)$ is analytic, and the Mellin-Parseval identity
\beq
\int_0^\ii f(x) g(x)\,dx=\frac1{2\pi i}
\int_{\Re z=c}\hat f(z)\hat g(1-z) \,dz\:,
\label{Mel-Par}
\eeq
valid for any pair of functions $f$, $g$
with Mellin transforms $\hat f$, $\hat g$.
Then choosing $f(t)=t^{-3/2}\Tr e^{-tL}$,
$g(t)=e^{-n^2\be^2/4t}$ and using Eqs.~(\ref{v3})
and (\ref{Mel-Par}), we obtain the so called Mellin-Barnes
representation for the statistical sum contribution
\cite{byts92-7-397}
\beq
\vone&=&-\frac{1}{2\pi i\vol}
\int_{\Re s=c}\Ga(s-1)\ze_R(s)
\ze(\fr{s-1}{2}|L)\;\be^{-s}\;ds\:.
\label{v33}
\eeq
This is an exact expression, which is particularly
useful in the derivation of the high temperature
expansion of the effective potential,
once the $\ze$-function related to the operator $L$
is known. In Eq.~(\ref{v33}), $\ze_R(s)$ is the usual Riemann
$\ze$-function.

\s{Selberg trace formula}
\label{S:STF}

The $\ze$-function of the Laplace-Beltrami operator
acting on the compact hyperbolic manifold $H^3/\Ga$ can be related
to the properties of the discrete group $\Gamma$ of
isometries of $H^3$ by means of Selberg trace formula.
To state such a relation, let $h(r)$ be an even and holomorphic
function in a strip larger than $2$ about the real axis and
$h(r)={\cal O}(r^{-(3+\ep)})$ uniformly in this strip
as $r\to\infty$. Then the Selberg trace formula reads
\cite{EGM87,selberg56}
(here we are considering untwisted scalar fields, namely functions such
that $\phi(\ga x)=\phi(x)$, $\forall\ga\in\Ga$)
\beq
\sum_{j=0}^{\infty} h(r_j)=\frac{V_{\cal F}}{2\pi^2}
\int_0^{\infty}r^2h(r)\,dr+\sum_{{\cal P}}
\sum_{k=1}^{\infty}\frac{l_\ga\hat h(kl_{\gamma})}
{S_3(k;l_{\gamma})}
\:.\label{STF}
\eeq
As usual, we normalize the constant curvature to $\ka=R/6=-1$.
In this manner all quantities will be dimensionless.
Here $V_{\cal F}$ is the volume of the fundamental domain ${\cal F}$,
which, in normal units, is just the volume of $H^3/\Ga$,
ans $\hat h$ is the Fourier transform of $h$, that is
\beq
\hat h(u)=\frac1{2\pi}\int_{-\infty}^{\infty}
e^{-iru}h(r)\,dr\:.
\eeq
To help the reader understanding the formula, we briefly explain the
double series on the right hand side. The symbol $\ca P$ denotes the set
of all primitive closed geodesics in the compact manifold, namely, those
geodesics passing only once through any given point. If $\ga\in\ca P$ then
its length is denoted by $l{\ga}$.
The integer $k$ is the winding number of any non primitive geodesic. Hence,
the sum over all primitive geodesics followed by the sum over $k$, which
appears in the trace formula, is nothing but the sum over all geodesics
of the compact manifold. This is expected because every geodesic must
contribute to the trace. Now to every closed primitive geodesic $\ga$,
with length $l{\ga}$, we can associate an isometry, also denoted
$\ga$, namely, the unique isometry such that
the hyperbolic distance $d(x,\ga x)=l_{\ga}$. Here $x\in H^3$
is any point whose projection in the quotient manifold belongs to $\ga$.
Thus $\ga\in PSL(2,\C)$ (the isometry group of $H^3$)
and so it can be conjugated to a diagonal matrix,
whose upper entry we call $a_{\ga}$ (the lower entry is then
$a_{\ga}^{-1}$). A simple calculation will shows that $l_{\ga}=\ln(|a_{\ga}|)$
and finally $S_3(k;l_{\ga})=|a_{\ga}^{k}-a_{\ga}^{-k}|^2$.
This is seen to be essentially the Jacobian
determinant of the parallel transport around $\ga$
(see Ref.~\cite{chavel84} for details).

In Eq.~(\ref{STF}), the sum on the left hand side
is over the spectrum of the Laplace operator
with eigenvalues $-\la_j$, and
$r_j=\sqrt{\la_j-1}$, if $\la_j>1$, while
$r_j=i\sqrt{1-\la_j}$ if $\la_j<1$.
Finally, the integral is the contribution of
those geodesics which are contractible to
a point, i.e. it is the direct path contribution to the trace.

In order to compute $\Tr\exp(-tL)$ by Selberg trace formula,
we may choose $h(r)=\exp(-t[r^2+M^2])$,
whose Fourier-transform reads
\beq
\hat h(u)=\frac{Me^{-tM^2-u^2/4t}}
{\sqrt{4\pi t}}
\eeq
Here we have set
\beq
M^2=V''(\hat\phi)-\ka=m^2+\frac{\la\hat\phi^2}{2}
+\left(\xi-\frac16\right)R\, .
\eeq

By a simple calculation we obtain
\beq
\Tr e^{-tL}=\frac{V_{\cal F}e^{-tM^2}}{(4\pi t)^{3/2}}
+\frac{e^{-tM^2}}{(4\pi t)^{1/2}}
\sum_{{\cal P}}\sum_{k=1}^{\infty}
\frac{l_\ga e^{-k^2l_\ga^2/4t}}{S_3(k;l_{\gamma})}
\:,\label{}\eeq
\beq
\ze(s|L)&=&\frac{V_{\cal F}}{(4\pi)^{3/2}}
\frac{\Ga(s-3/2)M^{3-2s}}{\Ga(s)}\nn\\
&&+\sum_{{\cal P}}\sum_{k=1}^{\infty}
\frac{l_\ga}{S_3(k;l_{\gamma})}
\aq\frac{kl_\ga}{2M}\cq^{s-1/2}
\frac{K_{s-1/2}(Mkl_\ga)}{\pi^{1/2}\Ga(s)}
\:,\label{33a}
\eeq
while the effective potential reads
\beq
\vone&=&-\frac{M^4}{2\pi^2}
\sum_{n=1}^\ii\frac{K_2(n\be M)}{(n\be M)^2}\nn\\
&&-\frac{M^2}{\pi\vol}\sum_{n=1}^\ii
\sum_{{\cal P}}\sum_{k=1}^{\infty}
\frac{l_\ga K_1\left(M\sqrt{n^2\be^2+k^2l_\ga^2}\right)}
{S_3(k;l_{\gamma})M\sqrt{n^2\be^2+k^2l_\ga^2}}
\:,\label{EP}
\eeq
where $K_\nu(s)$ represents the modified Bessel function.
{}From Eq.~(\ref{33a}) we immediately get the residue of $\ze(s|L)$ in
the poles $s=3/2-k$ ($k=0,1,2\dots$). That is
\beq
\Res(\ze(s|L);s=3/2-k)
=\frac{(-1)^kM^{2k}}{(4\pi)^{3/2}k!}
\frac{V_{\cal F}}{\Ga(3/2-k)}
\:.\label{ZFpole}
\eeq

To complete this section we introduce the Selberg $\Xi$ and
$Z$ funtions, which contain all topological informations.
They are defined by means of formula
\beq
\Xi(s)=\frac{d}{ds}\ln Z(s)=
\sum_{{\cal P}}\sum_{k=1}^{\infty}
\frac{l_\ga e^{-(s-1)kl_\ga}}{S_3(k;l_{\gamma})}
\:,\label{}\eeq
and can be found for example in Ref.~\cite{EGM87}
to which we ones more refer the reader for more details.
Since the number of
closed primitive geodesics with a given length $l_{\ga}$ is asymptotically
$l_{\ga}^{-1}\exp(2l_{\ga})$ and $S_3(k;l_{\ga})\simeq\exp(kl_{\ga})$, the
series defining the $\Xi(s)$-function is convergent for
$\Re s>2$. The Selberg function $Z(s)$,
on the other hand, is entire of order $3$ and has
zeroes at the points $s_j=1\pm ir_j$, whose order is the eigenvalues
multiplicity.

By means of the above functions, it is possible to rewrite
the topological contribution to $\ze(s|L)$. In fact we obtain
\beq
\ze(s|L)&=&\frac{V_{\cal F}}{(4\pi)^{3/2}}
\frac{\Ga(s-3/2)M^{3-2s}}{\Ga(s)}\nn\\
&&+\frac{\sin\pi s\,M^{1-2s}}{\pi}
\int_1^\ii
\Xi(1+Mu)\,(u^2-1)^{-s}\,du
\:,\label{ZF-chi}
\eeq
from which easily follows
\beq
\ze'(0|L)=\frac{V_{\cal F}M^3}{6\pi}-\ln Z(M+1)\:.
\eeq

Now, replacing $M^2$ with $M^2+t^2$ in the latter formula and using
Eq.~(\ref{v2}), we obtain the nice representation
\beq
\vone&=&-\frac{M^4}{2\pi^2}
\sum_{n=1}^\ii\frac{K_2(n\be M)}{(n\be M)^2}\nn\\
&&+\frac{1}{\pi\vol}\sum_{n=1}^{\ii}
\int_{0}^{\ii}\ln Z\at1+\sqrt{[t^2+M^2]/|\ka|}\ct\,
\cos n\be t\,dt
\:,\label{EP2}
\eeq
where normal units have been restored.

\s{Low and high temperature expansions}

The low temperature behaviour of the theory
can be directly obtained from  Eq.~(\ref{EP}) since the
asymptotic expansion of modified Bessel function is well known.
In fact we have
\beq
\vone&\sim&
-\frac{1}{2\vol}\frac{M^2}{(2\pi)^{1/2}}
\sum_{n=1}^\ii\sum_{{\cal P}}\sum_{k=1}^{\infty}
\frac{l_\ga e^{-n\be M\sqrt{1+(kl_\ga/n\be)^2}}}
{S_3(k;l_{\gamma})\left(n\be M\sqrt{1+(kl_\ga/n\be)^2}\right)^{3/2}}
\nn\\
&&\hs\X\aq 1+\frac3{4\left(n\be M\sqrt{1+(kl_\ga/n\be)^2}\right)}
\cq
-\frac12\frac{M^4}{(2\pi)^{3/2}}
\sum_{n=1}^\ii\frac{e^{-n\be M}}{(n\be M)^{5/2}}
\:.\label{LTE}
\eeq
We see that the leading term comes from the topological part.

In order to get the high temperature expansion of the effective
potential, we shall perform the integration in Eq.~(\ref{v33}).
To this aim we observe that the integrand function
in Eq.~(\ref{v33})
has simple poles at the points $s=4,2,1,-2k$
($k=1,2,\dots$) and a double pole at $s=0$.
Hence integrating this function along a path
enclosing all the poles, we obtain a series expansion in powers of
$\be$, whose coefficients are related to the
residues of $\ze$-function, which have been computed in Sec.~\ref{S:STF}
(see also Ref.~\cite{cogn92-7-3677}).
After a straightforward computation, using Eq.~(\ref{ZF-chi}) we obtain
the high temperature expansion
(here $\ga$ is the Euler-Mascheroni constant)
\beq
\vone&\sim&-\frac{\pi^2T^4}{90}
+\frac{M^2T^2}{24}-\aq\frac{M^3}{12\pi}-\frac{M}{2\vol}\ln Z(1+M)
\cq T\nn\\
&&-\frac{M^4}{32\pi^2}
\left[\ln\frac{M}{4\pi T}+\ga-\frac{3}{4}\right]\label{HTE} \\
&&-\frac{M^2}{2\pi\vol|\ka|^{1/2}}
\int_1^\ii\Xi(1+tM|\ka|^{-1/2})\,\sqrt{t^2-1}\,dt
+O(T^{-2})\nn
\:,
\end{eqnarray}
which in the case of flat, topological trivial space,
is in agreement with results of Ref.~\cite{habe82-23-1852}.

It is interesting to note that all coefficients
of the negative powers of $T$ are proportional to
the residues of $\ze(s|L)$ and so they do not depend on the topology.
Moreover, the topological term
independent of $T$ in the latter formula is the same, but the sign,
as the topological contribution to the zero temperature effective
potential, also after renormalization
(see Ref.~\cite{CKZ93}).
This means that topology enters the high temperature
expansion of the effective potential only with a term proportional to $T$.

\s{Phase transitions of the system}

The relevant quantity for analyzing the phase transitions of the system
is the mass of the field. The quantum corrections to the mass
are defined by means of equation
\beq
V^{(1)}(\hphi)=\Lambda_{eff}
+\frac12(m_0^2+m_T^2)\hphi^2
+{\cal O}(\hphi^4)\:,\label{34}
\eeq
where $\Lambda_{eff}$ (the cosmological constant) in general represents a
complicated expression not depending on the background field $\hphi$
while $m_0$ and $m_T$ represent the zero and finite temperature
quantum corrections to the mass $m$.
As has been shown in Ref.~\cite{CKZ93}, $m_0$
has curvature and topological contributions,
which help to break the symmetry (for $\xi<1/6$). On the contrary,
here we shall see that $m_T$ always helps to restore the symmetry.

By evaluating the second derivative with respect to $\hphi$ at the point
$\hphi=0$ of Eqs.~(\ref{v4}) and (\ref{EP2}),
we obtain for $m_T$ the two equations
\beq
m_T^2&=&\frac{\la}{2\vol}
\Tr\frac{e^{-\be Q_0}}{Q_0(1-e^{-\be Q_0})}
\:,\label{mT0}\\
&=&\frac{\la M_0^2}{4\pi^2}
\sum_{n=1}^\ii\frac{K_1(n\be M_0)}{n\be M_0}\nn\\
&&+\frac{\la}{2\pi\vol|\ka|}\sum_{n=1}^\ii
\int_0^\ii\frac{\Xi(1+\sqrt{[t^2+M^2_0]/|\ka|})}
{\sqrt{[t^2+M^2_0]/|\ka|}}\cos n\be t\,dt
\:,\label{mT}
\eeq
where $Q_0=Q_{\phi=0}=|-\lap_S+m^2+\xi R|^{1/2}$ and
$M_0=M_{\phi=0}=|m^2+(\xi-1/6)R|^{1/2}$.
{}From the exact formula, Eq.~(\ref{mT0}), we see that
the finite temperature quantum corrections to the mass are always positive,
their strength mainly depending on the smallest eigenvalues
of the operator $Q_0$.
This means that such a contribution always helps to restore the symmetry.
The second very interesting expression, Eq.~(\ref{mT}), gives the
mass in terms of geometry and topology of the manifold.
In fact, the $\Xi$-function is strictly related to the isometry group
$\Ga$, which realizes the non trivial topology of \M.

To go further, we compute the corrections to the mass in the low and
high temperature limits. Using Eqs.~(\ref{LTE}) and (\ref{HTE})
we obtain
\beq
m_T^2&\sim&\frac\la4\frac{M_0^2}{(2\pi)^{3/2}}
\sum_{n=1}^\ii\frac{e^{-n\be M_0}}{(n\be M_0)^{3/2}}\nn\\
&&+\frac{\la}{4(2\pi)^{1/2}\vol}
\sum_{n=1}^\ii\sum_{{\cal P}}\sum_{k=1}^{\infty}
\frac{l_\ga e^{-n\be M_0\sqrt{1+(kl_\ga/n\be)^2}}}
{S_3(k;l_{\gamma})\at n\be\sqrt{1+(kl_\ga/n\be)^2}\ct^{1/2}}
\label{LmT}\\
&&\hs\hs\X\aq 1+\frac1{4\at n\be M_0\sqrt{1+(kl_\ga/n\be)^2}\ct}
\cq\:,\nn
\eeq
\beq
m_T^2&\sim&\frac{\la T^2}{24}
-\frac{3M_0\la T}{8\pi}\nn\\
&&+\frac{|\ka|^{-1/2}}{4M_0^2\vol}
\aq\ln Z(1+M_0|\ka|^{-1/2})+M_0|\ka|^{-1/2}\,
\Xi(1+M_0|\ka|^{-1/2})\cq\,M_0\la T
\:,\label{HmT}
\eeq
which are valid for low and high temperature respectively.

\s{Conclusions}
In this paper we considered a self-interacting scalar field living on
an ultrastatic spacetime whose spatial part is a compact hyperbolic
manifold. We were especially interested in the finite temperature
effective potential of the theory, which serves as the effective
cosmological constant in new inflationary models.
As we have seen, the main technical tool to obtain the effective
potential was the Selberg trace formula, Eq.~(\ref{STF}).
Analytical expressions for the effective potential suitable in the
low, Eq.~(\ref{LTE}), and high temperature limit, Eq.~(\ref{HTE}),
have been obtained. Whereas at $T=0$ for $\xi<\frac16$ quantum
corrections to the classical potential can help to break symmetry
\cite{CKZ93}, the results show that, if the
temperature is high enough, quantum corrections always help to
restore the symmetry (see Eq.~(\ref{HmT})) .

\ack{K.~Kirsten is grateful to Prof.~M.~Toller, Prof.~R.~Ferrari and
Prof.~S.~Stringari for the hospitality in the Theoretical Group of the
Department of Physics of the University of Trento.}


\begin{thebibliography}{10}

\bibitem{BD82}
Birrell N and Davies~P~C W.
\newblock {\it Quantum Fields in Curved Spaces}.
\newblock Cambridge University Press, Cambridge, (1982).

\bibitem{brandenberger85}
Brandenberger~R H.
\newblock Rev.~Mod.~Phys., {\bf 57}, 1, (1985).

\bibitem{BOS92}
Buchbinder~I L, Odintsov~S D, and Shapiro~I L.
\newblock {\it Effective Action in Quantum Gravity}.
\newblock IOP Publishing, Bristol and Philadelphia, (1992).

\bibitem{AS82}
Albrecht A and Steinhardt~P J.
\newblock Phys.~Rev.~Lett., {\bf 48}, 1220, (1982).

\bibitem{L82}
Linde~A D.
\newblock Phys.~Lett.~B, {\bf 108}, 389, (1982).

\bibitem{CW73}
Coleman S and Weinberg~E J.
\newblock Phys.~Rev.~D, {\bf 7}, 1888, (1973).

\bibitem{oconnorhu84}
O'Connor~D J and Hu~B L.
\newblock Phys.~Rev.~D, {\bf 30}, 743, (1984).

\bibitem{guven87}
Guven J.
\newblock Phys.~Rev.~D, {\bf 35}, 2378, (1987).

\bibitem{CKV93unv}
Cognola G, Kirsten K, and Vanzo L.
Phys.~Rev.~D., {\bf 48}, 2813 (1993).

\bibitem{critchleyhustylianopoulos87}
Critchley R, Hu~B L, and Stylianopoulos A.
\newblock Phys.~Rev.~D, {\bf 35}, 510, (1987).

\bibitem{kirstenself}
Kirsten K.
Class.~Quantum Grav., {\bf 10}, 1461 (1993).

\bibitem{OHS83}
O'Connor~D J, Hu~B L, and Shen~T C.
\newblock Phys.~Lett.~B, {\bf 130}, 31, (1983).

\bibitem{OHS85}
O'Connor~D J, Hu~B L, and Shen~T C.
\newblock Phys.~Rev.~D, {\bf 31}, 2401, (1985).

\bibitem{oconnorhu86}
O'Connor~D J and Hu~B L.
\newblock Phys.~Rev.~D, {\bf 34}, 2535, (1986).

\bibitem{F84}
Futamase T.
\newblock Phys.~Rev.~D, {\bf 29}, 2783, (1984).

\bibitem{huang90}
Huang W-H.
\newblock Phys.~Rev.~D, {\bf 42}, 1287, (1990).

\bibitem{R87}
Ringwald A.
\newblock Ann.~Phys., {\bf 177}, 129, (1987).

\bibitem{R87a}
Ringwald A.
\newblock Phys.~Rev.~D, {\bf 36}, 2598, (1987).

\bibitem{hartlehu79}
Hartle~J B and Hu~B L.
\newblock Phys.~Rev.~D, {\bf 20}, 1772, (1979).

\bibitem{berkin92}
Berkin~A L.
\newblock Phys.~Rev.~D, {\bf 46}, 1551, (1992).

\bibitem{DJ74}
Dolan L and Jackiw R.
\newblock Phys.~Rev.~D, {\bf 9}, 3320, (1974).

\bibitem{A90a}
Actor A.
\newblock Class.~Quantum Grav., {\bf 7}, 1463, (1990).

\bibitem{FY79}
Ford~L H and Yoshimura T.
\newblock Phys.~Lett.~A, {\bf 70}, 89, (1979).

\bibitem{ford80}
Ford~L H.
\newblock Phys.~Rev.~D, {\bf 21}, 933, (1980).

\bibitem{T80}
Toms~D J.
\newblock Phys.~Rev.~D, {\bf 21}, 2805, (1980).

\bibitem{T80c}
Toms~D J.
\newblock Phys.~Rev.~D, {\bf 21}, 928, (1980).

\bibitem{KN81}
Kennedy G.
\newblock Phys.~Rev.~D, {\bf 23}, 2884, (1981).

\bibitem{W74}
Weinberg S.
\newblock Phys.~Rev.~D, {\bf 9}, 3357, (1974).

\bibitem{L79}
Linde~A D.
\newblock Rep.~Prog.~Phys., {\bf 42}, 389, (1979).

\bibitem{G82}
Goncharov~Y P.
\newblock Phys.~Lett.~A, {\bf 91}, 153, (1982).

\bibitem{DS81}
Denardo G and Spallucci E.
\newblock Nuovo Cim.~A, {\bf 64}, 27, (1981).

\bibitem{DS80}
Denardo G and Spallucci E.
\newblock Nuovo Cim.~A, {\bf 58}, 243, (1980).

\bibitem{ER90a}
Elizalde E and Romeo A.
\newblock Phys.~Lett.~B, {\bf 244}, 387, (1990).

\bibitem{EKK93unv}
Elizalde E and Kirsten K.
\newblock Topological symmetry breaking in self-interacting theories on
  toroidal spacetime.
\newblock Preprint Universit\`a di Trento, June 1993.

\bibitem{L83}
Linde~A D.
\newblock Phys.~Lett.~B, {\bf 123}, 185, (1983).

\bibitem{G84}
Goncharov~Y P.
\newblock Phys.~Lett.~B, {\bf 147}, 269, (1984).

\bibitem{ellis71}
Ellis~G~F R.
\newblock Gen.~Rel.~Grav., {\bf 2}, 7, (1971).

\bibitem{ellisschreiber86}
Ellis~G~F R and Schreiber G.
\newblock Phys.~Lett.~A, {\bf 115}, 97, (1986).

\bibitem{fagundes93}
Fagundes~H V.
\newblock Phys.~Rev.~Lett., {\bf 70}, 1579, (1993).

\bibitem{CKZ93}
Cognola G, Kirsten K, and Zerbini S.
\newblock Phys.~Rev.~D, {\bf 48}, 790, (1993).

\bibitem{BOS89}
Buchbinder~I L, Odintsov~S D, and Shapiro~I L.
\newblock Rivista Nuovo Cim., {\bf 12}, 1, (1989).

\bibitem{jackiw74}
Jackiw R.
\newblock Phys.~Rev.~D, {\bf 9}, 1686, (1974).

\bibitem{hawking77}
Hawking~S W.
\newblock Commun.~Math.~Phys., {\bf 55}, 133, (1977).

\bibitem{critchleydowker76}
Critchley R and Dowker~J S.
\newblock Phys.~Rev.~D, {\bf 13}, 3224, (1976).

\bibitem{byts92-7-397}
{Bytsenko A~A, Vanzo L and Zerbini S}.
\newblock Mod.~Phys.~Lett., {\bf {A7}}, {397}, (1992).

\bibitem{EGM87}
Elstrodt J, Grunewald F, and Mennicke J.
\newblock Math.~Ann., {\bf 277}, 655, (1987).

\bibitem{selberg56}
Selberg A.
\newblock J.~Indian Math.~Soc., {\bf 20}, 47, (1956).

\bibitem{chavel84}
Chavel F.
\newblock {\it Eigenvalues in Riemannian Geometry}.
\newblock Academic Press, New York, (1984).

\bibitem{cogn92-7-3677}
Cognola G and Vanzo L.
\newblock Mod.~Phys.~Lett., {\bf {A7}}, {3677}, (1992).

\bibitem{habe82-23-1852}
{Haber H E and Weldon H A}.
\newblock J.~Math.~Phys., {\bf {23}}, {1852}, (1982).

\end{thebibliography}
\end{document}